\documentclass[fleqn,10pt]{wlscirep}
\usepackage[utf8]{inputenc}
\usepackage[T1]{fontenc}
\usepackage{amsfonts}
\usepackage{amssymb}
\usepackage{amsmath}
\usepackage{calc}
\usepackage{graphicx}
\usepackage{bm}
\usepackage{verbatim}

\usepackage[normalem]{ulem}
\usepackage{amsmath,amssymb}
\usepackage{multirow}
\usepackage{xcolor,soul}
\usepackage{srcltx}
\usepackage{hyperref,graphicx}
\usepackage{ulem}

\renewcommand{\a}{\hat{a}}
\newcommand{\ac}{\hat{a}^\dagger}
\renewcommand{\b}{\hat{b}}
\newcommand{\bc}{\hat{b}^\dagger}
\newcommand{\sm}{\hat{\sigma}^-}
\renewcommand{\sp}{\hat{\sigma}^+}
\newcommand{\bra}[1]{\langle #1 |}
\newcommand{\ket}[1]{| #1 \rangle}

\newcommand{\eps}{\varepsilon}
	\title{Intermittent decoherence blockade in a chiral ring environment}

\author[1]{Salvatore Lorenzo}
\author[2]{Stefano Longhi}
\author[3]{Albert Cabot}
\author[3]{Roberta Zambrini}
\author[3,*]{Gian Luca Giorgi}
\affil[1]{Dipartimento di Fisica e Chimica, Universit\'a degli Studi di Palermo, via Archirafi 36, I-90123 Palermo, Italy}
\affil[2]{Dipartimento di Fisica, Politecnico di Milano, Piazza L. da Vinci 32, I-20133 Milano, Italy}
\affil[3]{IFISC (UIB-CSIC), Instituto de Fisica Interdisciplinar y Sistemas Complejos - Palma de Mallorca, E-07122. Spain}

\affil[*]{gianluca@ifisc.uib-csic.es}

\begin{abstract}
It has long been recognized that emission of radiation from atoms is not an intrinsic property of individual atoms themselves, but it is largely affected by the characteristics of the photonic environment and by the collective interaction among the atoms. 
A general belief is that preventing full decay and/or decoherence  requires the existence of dark states, i.e., dressed light-atom states that do not decay despite the dissipative environment. Here, we show that, contrary to such a common wisdom, decoherence suppression can be intermittently achieved on a limited time scale, without the need for any dark state,  when the atom is coupled to a chiral ring environment, leading to a highly non-exponential staircase decay. This effect, that we refer to as  \textit{intermittent decoherence blockade}, arises from periodic destructive interference  between light emitted in the present and light emitted in the past, i.e., from delayed coherent quantum feedback.
\end{abstract}
\begin{document}

\flushbottom
\maketitle
%
%
\thispagestyle{empty}


\section*{Introduction}

Spontaneous emission is a fundamental process in quantum optics and quantum electrodynamics \cite{Scu-Zu,Loudon,r1}.
	While in the most typical cases it is described by an exponential
	decay of a quantum   (atomic or solid state) system towards its ground state, accompanied
	by an irreversible emission of a photon \cite{r2}, the properties of the surrounding photonic environment \cite{purcell,itano,PC},
	as well as measurement  \cite{zeno}, or collective  effects \cite{r3,r3b}, 
	can largely affect spontaneous emission, with consequences ranging from control of single-photon sources to decoherence.

	Dimension and geometric constraints of the photonic environment (like cavities \cite{purcell}),  continuous or  discrete-mode structures of the reservoir \cite{nemet},  as well as engineered surrounding media (for instance exhibiting band-gaps \cite{PC}), can significantly enhance or inhibit the decay rate of a single emitter.
	Recently, 
	more complex photonic environments  have been shown to be powerful resources for controlling light-emitter interaction in unprecedented ways \cite{control,control2,control3,control4}.	
	
	Coupling one or more atoms to one-dimensional chiral waveguides or topological photonic structures, that break time reversal symmetry, enables to control 
	the directionality of spontaneous emission and to deeply modify photon-mediated interactions, with major applications in 
	the design of integrated non-reciprocal single-photon devices, spin-photon interfaces, and in the synthesis of novel
	quantum states such as entangled spin states and photonic clusters states  \cite{r7a,r7b,r7,r8,r9,r10,r11,r12}. Likewise,
	'giant' artificial atoms, in which the atomic dimension greatly exceeds the 'photon' wavelength and the time spent by light to cross the atom can not be neglected, 
	provide a new paradigm of atom-field interaction \cite{r13,r14,r15,r16,r17,r18,r19,r20,r20b,r21}. Since the atom cannot be considered point-like anymore, spontaneous emission
	ceases to be exponential and the decay dynamics is described by a differential-delayed equation \cite{r14,r16,r19,r21}, 
	displaying strictly non-Markovian (memory) effects arising from delayed coherent quantum feedback \cite{r22,r23,r24}.
	Similar memory-like effects  are also found in ordinary (point-like) atoms in the presence of mirrors or retardation effects \cite{r25,r26,r27,r28,r28a,r28b,r29,r30,r31,r31b}.
	
	One among the most striking phenomena achieved through complex environments engineering 
	is the possibility to inhibit spontaneous emission and dechoerence under certain geometric conditions, i.e. the stabilization of quantum superposition states in the presence of dissipation or other forms of decay channels or dephasing. This goal is of major relevance in different contexts ranging from quantum computation, where limiting effects of decoherence \cite{Qcompreview} and  decoherence-free-space have been broadly studied \cite{DFS_comp,DFS_comp2}, to     quantum biology \cite{r31c,r31d,r31e} and quantum chemistry \cite{r31f,r31g}, where pure dephasing and non-radiative channels  are the main sources that destroy electronic coherence in molecular dynamics.
	Such a decoherence/decay blockade stems from the appearance of  
	dressed light-matter states, commonly known as dark states, or else bound states in the continuum, that do not decay despite the dissipative environment.
	The existence of dark states and their ability to prevent quantum decay via 
	destructive interference among different decay channels has been known since long time and 
	studied in several areas of physics 
	\cite{r32,r33,r34,r35,r36,r37,r38,r39,r40,r41,r41b,r42,r43,r44,r45,r46,r47,r48,r49,r50a}, along 
	with the related concept of decoherence-free subspaces \cite{r50}, i.e.
	regions in Hilbert space which are not affected by decoherence.
	A fully open question  is whether spontaneous emission and decoherence 
	can be inhibited, at least transiently or intermittently, in the absence of any decoherence-free subspace,
	or even though the atom-light system does not show any dark state.\par 
	In this work we show rather surprisingly that, harnessing the idea of 
	delayed coherent quantum feedback in a reservoir with effective discrete and continuous mode structure, a point-like 
	atom emitting in a chiral ring photonic waveguide, sustaining slow and fast counter-propagating 
	photonic modes, undergoes intermittent decoherence suppression on a fast time scale, 
	displaying an exotic staircase decay dynamics. 
	Such an effect, that we refer to as  \textit{intermittent decoherence blockade}, 
	arises from periodic destructive interference between light emitted in the present, both in 
	fast and slow photonic modes, and light emitted in the past in the fast photonic modes. Due to the different group velocities of counter-propagating chiral modes in the ring, two different time scales are involved in the decay dynamics, which are determined by the energy level spacing in the slow and fast bands of the ring waveguide. On the fast time scale, which is of major interest in our work, the atom turns out to be 
	effectively coupled with both a continuous set of modes (the modes of the slow ring band) into which irreversible decay and decoherence occur, and a discrete set of modes (the modes of the fast ring band), which provide delayed feedback and re-coherence in the atom and are thus ultimately responsible for the intermittent decoherence blockade predicted in our work. Clearly, on the long time scale, determined by the energy spacing between the levels in the band with the slow group velocity, the full discrete nature of the reservoir will lead to a fully coherent and unitary dynamics, and the system will almost surely return to  its initial state according to the quantum recurrence theorem \cite{recur}.

\section*{Results}

\subsection*{Decoherence dynamics of an atom coupled to a chiral ring.}
	
\begin{figure}[t!]
		\begin{center}
			   \includegraphics[width=0.9\linewidth]{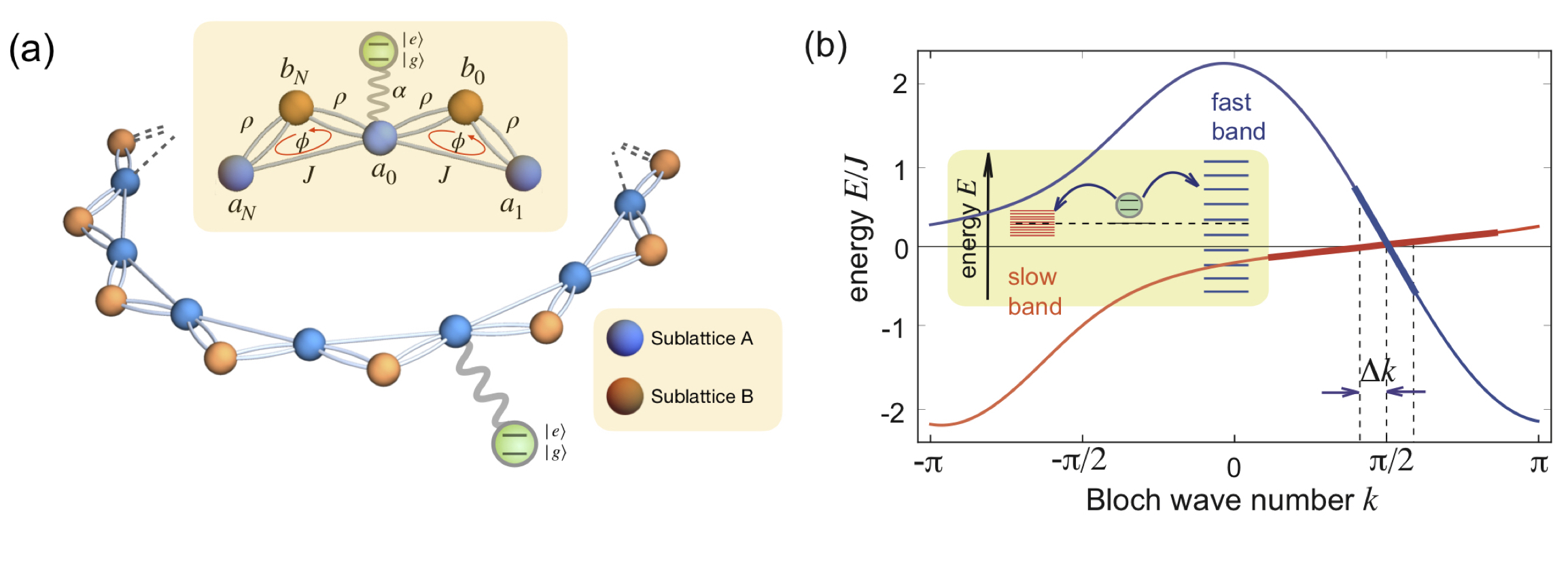}\\
			\end{center}
		\caption{(color online) (a) 
		Schematic of a two-level quantum emitter decaying on a chiral sawtooth photonic lattice. The lattice comprises two sublattices A and B, with hopping constants $J$ and $\rho$ (represented by single and double lines), on a ring geometry. A synthetic magnetic flux $\phi$ is applied in each plaquette of the  lattice (see the inset). (b) Energy diagram (dispersion curves) of the bipartite ring lattice for $\rho / J=0.5$ and $\phi=\pi/2$. Note the band closing point at $k= \pi /2$ due to the $\phi= \pi/2$ flux condition, and the existence of fast and slow bands crossing at the gap closing point, corresponding to counter-propagating modes with fast and slow group velocities $\Omega^{\pm}$. Quantization of the Bloch wave number $\Delta k = 2 \pi /N$ due to the ring boundary conditions introduces two energy scales (i.e. energy quantization) for fast and slow bands, into which the quantum emitter decays (see the inset).}\label{fig1}
	\end{figure}

We consider the decay/decoherence dynamics of a two-level atom
		coupled to the radiation modes of an engineered chiral bath with broken time reversal symmetry. 
		The photonic bath realizes a chiral sawtooth waveguide \cite{r10,r11}, consisting of a bipartite lattice of cavities/resonators composed by two sublattices A and B in a ring geometry, and threaded by a synthetic gauge field $\phi$ in each plaquette, as schematically depicted in Fig.\ref{fig1}(a). Such a model system has been investigated in some recent works and can be physically implemented in different 
		platforms, such as squids, cold atoms, and integrated photonic circuits \cite{r10,r11,r12}.
	The bath is governed by the nearest-neighbor tight-binding  Hamiltonian \cite{r11}
	\begin{eqnarray}
	\hat{H}_B=\sum_{n=1}^N\left\{ \omega\left(\ac_n\a_n{+}\bc_n\b_n\right){+}\left[\a_{n}\left(J\ac_{n+1}{+} 
	\rho e^{-i\phi}\bc_{n-1}{+}
	\rho \bc_{n}\right)+\text{h.c.}\right]\right\},
	\label{Hbath}
	\end{eqnarray}
	where $\a_n$ and $\b_n$ are the annihilation operators of the n-th $a$
	and $b$ modes with the same frequency $\omega$ (henceforth we will work in a frame rotating  at  $\omega$). 
    The  constants $J$ and $\rho$ are the nearest-neighbour coupling between the A lattice sites  and the hopping strength between
	the $a$ and $b$ modes respectively, as shown in Fig.\ref{fig1}(a). 
	This simple one-dimensional model admits complex couplings between lattice vertices, defining an effective magnetic flux per loop denoted by $\phi$ in Eq.(\ref{Hbath}).  As detailed  in Ref. \cite{r11}, the  saw-tooth lattice can be realized using superconducting qubits by coupling several single-loop plaquettes, experimentally implemented in Ref. \cite{martinis}. In such a scenario, typical lifetimes for the tight-binding Hamiltonian modes can be extracted from the experimental measurements of Refs. \cite{martinis,martinis2} and range from a few $\mu$s to a few tens of $\mu$s.  
	 Given the bipartite nature of the bath, the energy spectrum of $\hat{H}_B$ comprises two energy bands (Fig.\ref{fig1}(b)), with 
		dispersion relations given by (see Methods)
	\begin{eqnarray}
	E_k^{\pm}=J\cos k\pm\sqrt{J^2\cos^2k+2\rho^2(1+\cos(\phi+k))}.
	\label{bands}\end{eqnarray}
	 A non-vanishing magnetic flux $\phi$ breaks the time reversal symmetry of $\hat{H}_B$, i.e. $E_{-k}^{\pm} \neq E_{k}^{\pm}$. Remarkably, 
		for $\phi= \pi/2$ the intraband gap closes, giving raise to a band crossing at $k=\pi/2$ \cite{r11} with energy $E_{\pi/2}^{\pm}=0$.
		Near the band crossing point,
		the dispersion relations of the two bands show an almost linear behavior, resulting in a slow ($\Omega^+$) and fast ($\Omega^-$) group velocities 
		with opposite signs for the two bands, see fig.\ref{fig1}(b).
	
	A point-like emitter decays into the chiral bath via a weak coupling to the radiation mode of one resonator of sublattice A. Modelling the emitter as a two level system, $\{\ket{g},\ket{e}\}$, with energy separation $\omega_e$, we write the free Hamiltonian of the emitter and interaction Hamiltonian with the bath as
	\begin{eqnarray}
	\hat{H}_e+\hat{H}_{int}=\omega_e\ket{e}\bra{e}+\alpha\left(\sm\ac_0+\sp\a_0\right),
	\label{Hint}
	\end{eqnarray}
	where $\sigma^\pm$ are the usual ladder Pauli operators.
	As shown in  the Methods, the bath Hamiltonian can be diagonalized introducing slow and fast modes $s_k$ and $f_k$, 
	\begin{eqnarray}
	\hat{H}_B{=}\sum_{k}E^-_k\hat{s}^\dagger_k\hat{s}_k+E^+_k\hat{f}^\dagger_k\hat{f}_k.
	\end{eqnarray}
	In this representation the interaction part of equation\eqref{Hint} results to be
	\begin{eqnarray}
	H_{int}=\sum_{k} \frac{\alpha_k^-}{\sqrt{N}}\sp\hat{s}_k+\frac{\alpha_k^+}{\sqrt{N}}\sp\hat{f}_k + h.c.
	\label{Hint2}\end{eqnarray}
	with couplings $\alpha_k^\pm=\alpha E^\pm_k/N^\pm_k$, where $N^\pm_k=\sqrt{(E_k^{\pm })^2 -E_k^+ E_k^-}$.

	Supposing that the initial state is $\ket{\phi(0)}=\ket{e}{\otimes}\ket{vac}$ ($\ket{vac}$ denotes the vacuum state of the bath), it will evolve 
	into $\ket{\phi(t)}=\eps(t)\ket{e}{\otimes}\ket{vac}+\sum_{k,\pm}c_k^\pm(t)\ket{g}\otimes\ket{\psi_k^\pm}$, with $\eps(0)=1$. Following standard procedures (see the Methods), we arrive at the integral differential equation for the emitter excitation amplitude 
	\begin{eqnarray}
	\dot{\eps}(t)=-i\omega_e \eps(t)-\frac{1}{N}\int_{0}^{t}\!ds\;\eps(t-s) \sum_{k,\pm}\vert\alpha_k^\pm\vert^2 e^{-iE_k^\pm s}.
	\label{deps}
	\end{eqnarray}
	For $\phi{=}\pi/2$ the two dispersion curves \eqref{bands} can be linearized near $k=\pi/2$. Assuming the emitter resonant with modes near the crossing point ($\omega_e=0$, or $\omega_e=\omega$ in the lab frame) the equation for excitation amplitude \eqref{deps} becomes
	\begin{eqnarray}
	\dot{\eps}(t)\simeq-{\frac{1}{N}}\int_{0}^{t}\!ds\;\eps(t{-}s)\sum_{k,\pm}\vert\alpha_{\pi/2}^\pm\vert^2 e^{-i\Omega^\pm(k-\frac{\pi}{2})s}, \label{deps2}
	\end{eqnarray}
	where $\Omega^\pm{=-}J{\pm}\sqrt{J^2{+}\rho^2}$. Note that, according to Eqs.(\ref{deps}) and (\ref{deps2}), the amplitude $\epsilon(t)$ and its derivative $\dot{\epsilon}(t)$ are continuous functions of time $t$, as it should be on physical grounds, However, as discussed in the next section, for a large number $N$ of sites in the ring the numerically-computed evolution of $\epsilon(t)$ displays rather sharp changes at time instants corresponding to periodic feedback from the ring.  Such a scenario, which is similar to the dynamical behavior observed for a point-like two-level atom radiating in front of a mirror 
\cite{r25,r27,r28}, can be captured by approximating the exact integro-differential equations (\ref{deps},\ref{deps2}) with a differential-delayed equation \cite{r20b,r28}. To this aim, let us note that in the large $N$ limit the sum over $k$ on the right hand side of Eq.(\ref{deps2}) can be approximated by Dirac comb
 \small
	\begin{eqnarray}
	\sum_{k}e^{-i\Omega^\pm k s}{\sim} e^{-i\frac{N{-}1}{N}\pi\Omega^\pm s}\frac{N}{\vert\Omega^\pm\vert}\sum_{n=0}^{\infty}(-1)^{n(N+1)}\delta\left(s-\frac{nN}{\vert\Omega^\pm\vert}\right).
	\nonumber\label{diraccomb}
	\end{eqnarray}
\normalsize
	In this way the time integration in \eqref{deps2} can be performed leading to 
	\begin{eqnarray}
	\dot{\eps}(t){=}-\tfrac{1}{2}\gamma_0\eps(t)-\sum_{n=1}^{\infty}\sum_{\pm}\gamma_n^\pm
	\eps\!\left(t{-}nT^\pm\right)\Theta\!\left(t{-}nT^\pm\right), \label{deps3}
	\end{eqnarray}
	where 
	\begin{equation}
	\gamma_n^\pm{=}{ \frac{\vert\alpha_{\pi/2}^\pm\vert^2 }{\vert\Omega^\pm\vert}  e^{\pm i\pi n\tfrac{N}{2}}},\quad T^\pm=\frac{N}{\vert\Omega^\pm\vert},
	\label{gamma_n}\end{equation}
	 $\gamma_0{=}\gamma_0^+{+}\gamma_0^-$,
	and $\Theta(x)$ stands for the Heaviside step function.
		From inspection of equation (\ref{gamma_n}) we anticipate that, unless $N=2(2M)$, the damping rates exhibit a peculiar alternate change of sign at any round, increasing $n$.
	Note that by solving equation (\ref{deps3}), the behavior of coherences is also known. In fact, the reduced density matrix of the two-level atom at time $t$ is given by
	\begin{equation}
	{\rho}(t)=
	\begin{pmatrix}|\eps(t)|^2\rho_{ee}&\eps(t)\rho_{eg}\\
	\eps(t)^*\rho_{ge}&(1{-}|\eps(t)|^2)\rho_{ee}{+}\rho_{gg}\end{pmatrix}\,,
	\label{map}
	\end{equation}
	where $\rho_{jk} =\langle j|\rho(0)|k\rangle$ with $j,k=g,e$ are the entries of the possible mixed initial atomic density matrix $\rho(0)$.
	The first term in the right hand side of \eqref{deps3} represent the initial decay into ``both''  channels,  while the successive terms take into account back excitation from the bath into the emitter when the slow and fast modes make entire loops through the ring, i.e. in the presence of delayed coherent quantum feedback. The decay dynamics is thus governed by three different time scales: (i) the decay time $T_d=1/\gamma_0$, i.e. the inverse of the decay rate $\gamma_0=\gamma_0^++\gamma_0^-$ as determined by the usual Fermi golden rule in the weak-coupling and $N \rightarrow \infty$ limits or, equivalently, by the spectral density, that is, by the Fourier transform of the sum in equation (\ref{deps}), in a master equation approach; this term is proportional to $\alpha^{-2}$; (ii) and (iii) the feedback delay times $T^{\pm}=N / | \Omega^{\pm}|$ for the fast and slow decay channels, depending only on bath parameters. Note that, in the limit $ \rho / J \rightarrow 0$, the slow decay channel corresponds to a vanishing group velocity $\Omega^+ \rightarrow 0$ (flat band limit), i.e. in an extremely long delay feedback time $T^+$. Here we restrict our analysis considering the decay dynamics on a time scale shorter than $T^+$, so that the slow decay channel (slow band) can be regarded as a true bath with continuous energy spectrum, into which the point emitter continuously decays. 
	For times longer than $\sim T^-$, the discreteness of energy levels in the fast band cannot be neglected. So, on the one hand, the atom decays into a very continuum of modes (slow band), while, on the other hand, it is periodically fed by  the discrete modes of the fast band, which are responsible for delayed coherent quantum feedback. For a detailed discussion about delayed quantum feedback induced by reservoirs with continuous and discrete modes see Ref. \cite{nemet}.

	\begin{figure}[t!]
	\begin{center}
       \includegraphics[width=0.95\linewidth]{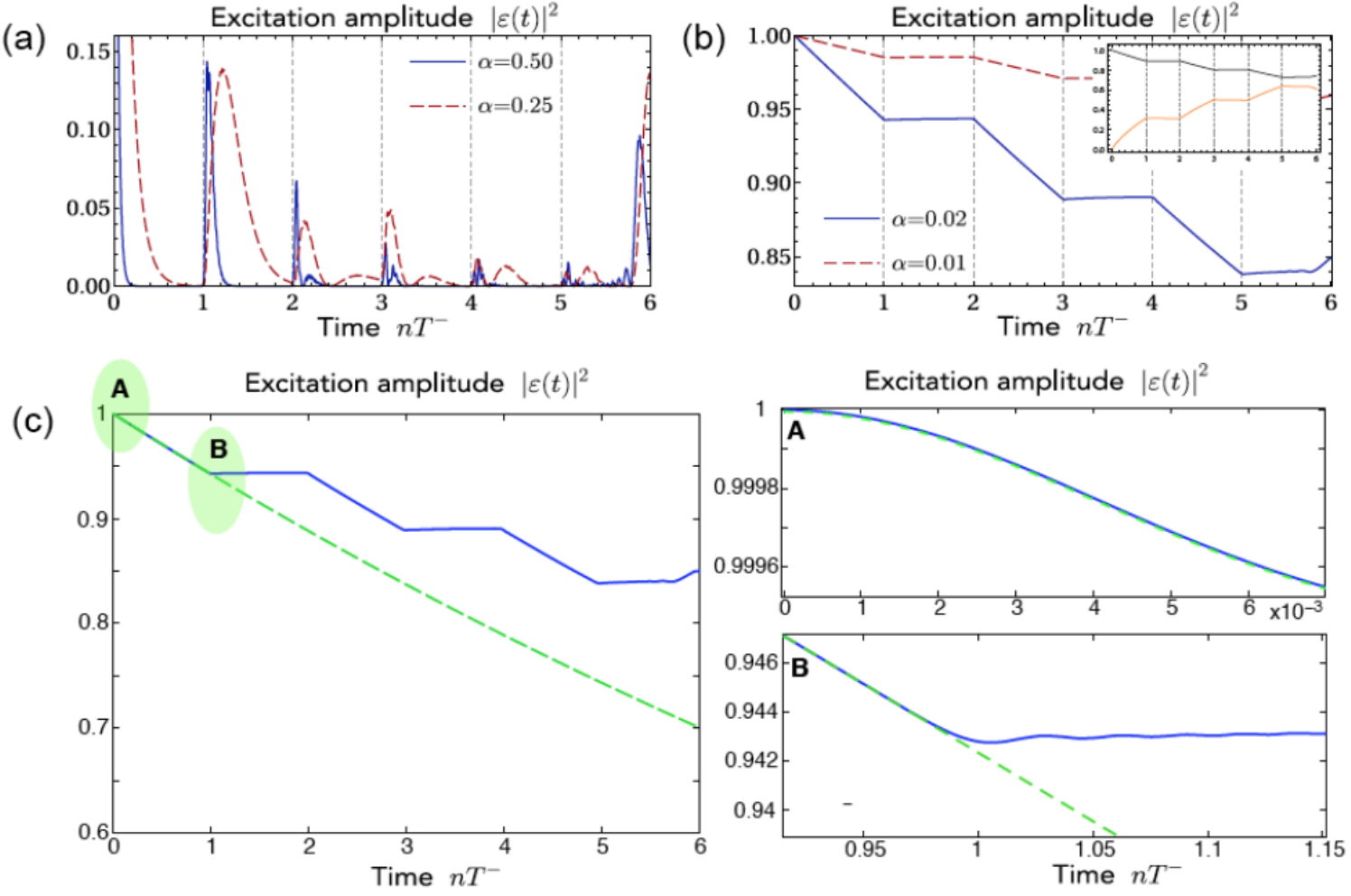}
	\end{center}
	\caption{(color online) (a,b) Evolution of the excitation amplitude $|\varepsilon(t)|^2$, solution of equation \eqref{deps} for $N=502$, $J=\rho=1$ and different values of $\alpha$. (a) ($\alpha=0.25$ and $\alpha=0.50$) In this regime a non exponential decay dynamics is observed, in which multiple revivals occur for $t=nT^-$. (b) ($\alpha=0.01$ and $\alpha=0.02$) In this regime intermittent blockade of the decay is observed until the time $t \sim 6 T^-$; at longer times feedback from the slow modes arises, which breaks intermittent decay suppression. Inset: evolution of purity (black) and von Neumann entropy (orange) for $\alpha=0.02$. (c) Comparison between decay dynamics for $\alpha=0.02$, $N=502$ [solid curve, as in panel (b)] and for $\alpha=0.02$, $N=\infty$ (dashed curve, continuous limit without delayed feedback). Note that for $0<t<T^-$ the two curves are overlapped. The plots on the right side in A and B depict an enlargement of the decay dynamics near the early time $t=0$, where the decay is parabolic (Zeno dynamics, panel A), and near the time $t=T^-$, where stopping of decoherence decay is observed in the finite $N$ case due to coherent delayed feedback.
	}\label{fig2}
	\end{figure}
	
	\begin{figure*}[t!]
	    \includegraphics[width=\linewidth]{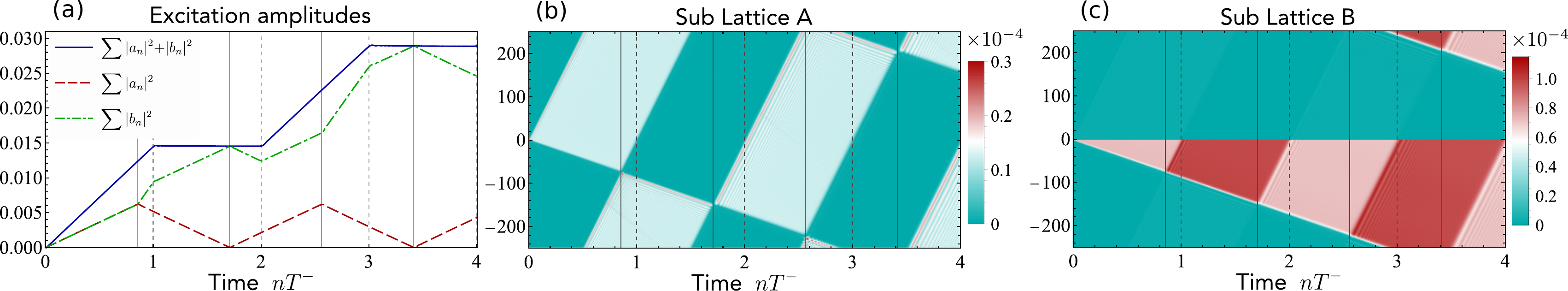}\\
	\caption{(color online) (a) Temporal evolution of the total excitation amplitude relative to sublattices A and B for in the intermittent decay blockade regime ($N=502$, $J{=}\rho{=}1$ and $\alpha=0.01$). (b),(c) Corresponding detailed evolution of the excitation amplitudes in sublattices A and B on a pseudocolor map (see the Methods for the definition of such amplitudes) . In each plot, vertical lines correspond to  times (solid) at which slow and fast waves meet and to times (dotted) at which the fast one makes a complete loop.
	}
	\label{fig3}
	\end{figure*}

\subsection*{Intermittent decoherence blockade in a chiral bath: delayed coherent quantum feedback and staircase dynamics} 
According to our theoretical analysis, two very distinct dynamical scenarios are expected depending on whether $T_d \ll T^-$ or $T_d \gg T^-$. While in the former case the decay dynamics shows typical multiple revivals, as observed e.g. in giant atom decay dynamics \cite{r16}, in the latter case 
a fully distinct behavior is observed, characterized by intermittent decays separated by intervals of decay suppression, corresponding to intermittent decoherence blockade. We computed the exact decay dynamics of $\epsilon(t)$ by numerically solving the coupled equations (\ref{exact}) given in the Methods, which are equivalent to the integro-differential equation (\ref{deps}).

 An example of multiple revivals in the decoherence dynamics is shown in Fig.\ref{fig2}(a), where the numerically-computed solution of \eqref{deps} is reported for $T_d \ll T^-$ and $T^+=2\sqrt{2} +3$.
Note that, after each time interval of duration $T^-$, excitation almost completely decays into the bath, however incomplete recoherence  is periodically observed due quantum feedback from the fast channel into the atom. On other hand when $T_d \gg T^-$, i.e. $\alpha\ll 1$, a surprising result is observed provided that $N=2(2s+1)$  with $s$ integer, namely decoherence can be intermittently suppressed (see  the Methods for the dynamics with different values of  $N$). An example of the decoherence dynamics in this regime obtained by simulation of the exact dynamics, equation (\ref{deps})  is shown in Fig. \ref{fig2}(b,c) and Fig. \ref{fig3}(a). Clearly, the decay of coherence $|\epsilon(t)|$ largely deviates from an exponential decay during time $T_d$ and, most importantly, shows a nearly staircase behavior, where decoherence is inhibited at alternating time intervals of duration $T^-$, while it displays an almost linear decay outside these intervals.
In the inset of Fig. \ref{fig2}(b), we show that this peculiar behavior can be found looking at basis-independent indicators, such as the purity (defined as the trace of $\rho^2(t)$) and the von Neumann entropy $S(\rho(t))=-{\rm Tr}[\rho(t) \log \rho(t)]$.  Figure \ref{fig2}(c) compares the intermittent decoherence blockade dynamics, arising from delayed quantum feedback of the ring bath geometry, with the conventional nearly-exponential decay dynamics that one would observe in the continuous limit $N=\infty$, where the feedback is absent. For times $t<T^-$, the decay dynamics in the two cases is clearly the same. In particular, in the very early stage near $t=0$, the decay is parabolic (Zeno interval, panel A), as it should be. Likewise, in the early stage the purity dynamics shows a parabolic (Gaussian) behavior, which is not however visible on the time scale depicted in the inset of Fig. 2(b).  In the continuous limit $N=\infty$, after the Zeno time the decay becomes exponential with a decay rate that turns out to be in  excellent agreement with the Fermi golden rule prediction $\gamma_0$. Note that, since $\gamma_0 T^- \ll 1$, the exponential decay in the range $(0,T^-)$ is very close to a linear decay. The dynamical behavior near the time $t=T^-$ is shown in panel B. As one can see, while in the continuous limit the decay is not interrupted (absence of delayed feedback from the bath), destructive interference leading to almost stopping of the decay is observed for finite $N$.

In order to better understand this
 intermittent decoherence blockade, one should look at the solution to (\ref{deps3}), which can be given in terms of  confluent hypergeometric functions, as shown in the Methods. In particular, the Taylor expansion of $\eps(t)$ up to the second order in $\alpha$, as obtained from the solution to equation (\ref{deps3}), reads explicitly
\small
\begin{eqnarray}\eps(t){\sim}\begin{cases}
 1{-}\tfrac{1}{2}\gamma_0(t{-}n T^-)\;\;\;\; \text{for}\qquad(2n)T^-{\leq}\;t\;{\leq} (2n{+}1)T^-\nonumber\\
1{-}\tfrac{1}{2}\gamma_0 (n+1) T^-\;\;\text{for}\;\;(2n{+}1)T^-{\leq}\;t\;{\leq} (2n{+}2)T^-,\nonumber
\end{cases}
\end{eqnarray}
\normalsize
with $n=\{0,1,2,3,\hdots\}$ such that $nT^-<T^+$.
It is evident now, how the coherence decay results intermittently suppressed, for an interval of time equal to $T^-$. The intermittent decoherence blockade can be traced back to destructive interference 
between light emitted in the past in the fast band and light emitted in the present in both fast and slow bands of the bath, as shown in Fig. \ref{fig3}. The figure depicts the total excitation present in sublattices $A$ and $B$. In a counter intuitive way, we see that every time the "two waves", slow and fast, meet [this happen at times $t^*{=}n T^-T^+/(T^-+T^+)$, vertical solid lines in Fig.\ref{fig3}], the excitation accumulated on sublattice $A$ is transferred to sublattice $B$, while the overall decay of the emitter is blocked at different alternating intervals (vertical dotted lines in Fig. \ref{fig3}). We emphasize that such an intermittent suppression of decoherence arising from delayed coherent quantum feedback does not require the existence of dark states, contrary to other decoherence suppression methods based on delayed feedback \cite{r22,r30,r31}. Indeed,   as shown in the Methods, the full Hamiltonian 
$\hat{H}=\hat{H}_{B}+\hat{H}_{e}+\hat{H}_{int}$ does not sustain eigenmodes localized near the emitter.  

Finally, it should be mentioned  that a necessary condition for the observation of the intermittent (staircase) decoherence behavior is the resonance between the frequency of the emitter and the  band crossing energy of the bath, which allowed us to perform the linearization of equation (\ref{deps2}). This resonance would be lost if for instance we changed the interaction Hamiltonian, defined by Eq.(\ref{Hint}), by considering a pure dephasing coupling of the form $\hat{H}_{\rm dep}=\alpha \hat\sigma_z(\hat a_0+\hat a_0^\dag)$. In this case, the decoherence caused by the bath (only observable if  the non-diagonal elements of the emitter density matrix were populated at time $t=0$) would be proportional to the scalar product  $\bra{vac}e^{-i \hat (\hat{H}_{B}+\hat{H}_{\rm dep}^\prime) t}\ket{vac}$ (where $\hat{H}_{\rm dep}^\prime=\alpha (\hat a_0+\hat a_0^\dag)$ is obtained from replacing $\hat\sigma_z$ with its eigenvalue $1$ over the excited state of the emitter in $\hat{H}_{\rm dep}$) which would depend on the whole spectrum and would not display any sign of resonance.

\section*{Discussion}


Quantum emitters coupled to  photonic waveguides represent a powerful integrated platform for  complex quantum  networks. While in typical scenarios the interaction between atoms and light can be treated using a master equation obeying the Born-Markov approximation, nontrivial phenomena can emerge beyond this limit. In particular, the Markovian approximation clearly fails in the presence of delayed coherent
quantum feedback, where one has to take into account the effects of the finite propagation speed of light, which introduces an effective memory.

The possibility to block decoherence is generally associated with the presence of frequency gap environments or, through a less trivial mechanism, due to the presence of bound states into the continuum. 
 In this work we showed a different scenario that exploits the chirality of the environment,   which is responsible, at the same time, for the decay into a continuum of modes and delayed feedback due to a discrete set of modes. Indeed, the interference effects of delayed coherent quantum feedback  enables intermittent  decoherence  suppression. Our results go beyond previous findings, which linked  the presence of coherent quantum feedback to the emergence of dressed light-atom dark states that cause light to be trapped around the emitter, as for instance in the famous atom-in-front-of-a-mirror  example. We showed that chiral waveguides can be exploited to generate nontrivial interference patterns, as they are able to separate fast and slow wavepackets, altering in this way the effect of  delayed feedback. The surprising result is the total suppression of decoherence during  finite time windows (whose length can be tailored modifying the system parameters) even in the absence of dark states. 
 As an additional comment, we point out that, since the photonic environment can be readily reconfigured by tuning the gauge phase $\phi$, we can open a wide gap and fully stop the decay of the quantum emitter on demand. Moreover, since the intermittent decoherence blockade is ultimately an interference effect which is very sensitive to the gauge phase $\phi$, besides controlling decoherence our setup could be of relevance for quantum sensing. To summarize,
  our results suggest that chiral waveguides together with delayed coherent
quantum feedback  represent a powerful reservoir
engineering tool of potential relevance in quantum manipulation and control, quantum sensing and quantum networks.

\section*{Methods}
\subsection*{$\hat{H}_B$ diagonalization}
Under periodic boundary conditions,  we can introduce momentum operators, $a_k(\b_k)=1/\sqrt{N}\sum_{n=1}^Ne^{-ikn}\a_n(\b_n)$, that allow us, moving to the rotating frame at the bare energy $\omega$, to rewrite the 
Hamiltonian $\hat{H}_B$ in the following form
\begin{eqnarray}
\hat{H}_B=\sum_k \begin{pmatrix}\ac_k&\bc_k\end{pmatrix}h_k\begin{pmatrix}\a_k\\\b_k\end{pmatrix},
\end{eqnarray}
where
\begin{eqnarray}
h_k=\begin{pmatrix}J(e^{-ik}+e^{ik})&g_k^*\\
g_k& 0 \end{pmatrix},
\label{hk}\end{eqnarray}
with $g_k=\rho(1+e^{-i(\phi+k)})$.
Each $h_k$ can be diagonalized with its respective unitary matrix 
\begin{eqnarray}
U_k=\begin{pmatrix}
E^-_k/N^-_k& E^+_k/N^+_k\\
g_k/N^-_k& g_k/N^+_k
\end{pmatrix},
\end{eqnarray}
with $N_k^\pm=\sqrt{(E^\pm_k)^2+\vert g_k\vert^2}$.
In this way, the slow and fast modes $s_k$ and $f_k$ are defined as
\begin{eqnarray}
\begin{pmatrix}\hat{s}_k\\\hat{f}_k\end{pmatrix}=U\begin{pmatrix}\a_k\\\b_k\end{pmatrix}=\begin{pmatrix}\frac{E^-_k}{N^-_k}\a_k+\frac{E^+_k}{N^+_k}\b_k\\\frac{g_k}{N^-_k}\a_k+\frac{g_k}{N^+_k}\b_k\end{pmatrix}.
\end{eqnarray}
\subsection*{Derivation of equation for emitter excitation amplitude}

An initial state of the form
$\ket{\phi(0)}{=}\ket{e}{\otimes}\ket{vac}$
will evolve in time as
\begin{equation}
\ket{\phi(t)}=\eps(t)\ket{e}{\otimes}\ket{vac}+\sum_{k,\pm}c_k^\pm(t)\ket{g}\otimes\ket{\psi_k^\pm}\nonumber,
\end{equation}
where $\eps(0)=1$, $c_k^{\pm}(0)=0$, and where $ \ket {\psi_k^+}=\hat{s}_k^\dag | vac \rangle$, $ \ket{ \psi_k^-}=\hat{f}_k^\dag | vac \rangle$.
The time-dependent Schr\"{o}dinger equation yields the set of coupled differential equations
\begin{eqnarray}
i\dot{\eps}(t)&=&\omega_e\eps(t)+\frac{1}{\sqrt{N}}\sum_{k\pm}\alpha_k^\pm c_k^\pm(t),\nonumber \\
i\dot{c}_k^\pm(t)&=&E_k^\pm c_k^\pm(t)+\frac{1}{\sqrt{N}}\alpha_k^\pm \eps(t). \label{exact}
\end{eqnarray}
Formally integrating the latter and substituting in the former, one arrive at the integral differential equation for the emitter excitation amplitude $\eps(t)$:
\begin{equation}
\dot{\eps}(t)=-i\omega_e \eps(t)-\frac{1}{N}\int_{0}^{t}ds \sum_{k\pm}\vert\alpha_k^\pm\vert^2 \eps(s)e^{-iE_k^\pm (t-s)}.
\end{equation}

\subsection*{Absence of dark states}
In the single-excitation sector, the energy spectrum $E$ and corresponding eigenstates of the full Hamiltonian $\hat{H}=\hat{H}_{B}+\hat{H}_{e}+\hat{H}_{int}$ are obtained by solving the linear eigenvalue problem
\begin{eqnarray}
E a_n & = & J(a_{n+1}+a_{n-1})+ \rho b_n + \rho \exp(-i \phi) b_{n-1} \nonumber \\
& + & \alpha \epsilon \delta_{n,1} \nonumber \\
E b_n & = &  \rho a_{n}+ \rho \exp(i \phi) a_{n-1} \label{matrice} \\
E \epsilon & = & \omega_e \epsilon+ \alpha a_1, \nonumber
\end{eqnarray}
($n=1,2,...,N$) where $\epsilon$, $a_n$ and $b_n$ are the excitation amplitudes in the atom and in the sublattices A and B, respectively. A dark state corresponds to a localized state near the emitter, the degree of localization being measured by the participation ratio $PR= (\sum_l |c_l|^2)^2 / \sum_l |c_l|^4$, where $\{c_l\}=\{a_n, b_n, \epsilon\}$ and $l=1,2,...,2N+1$. For localized
modes, $PR \sim 1$ while for extended states $PR \sim N$. Let us assume the atom in resonance with the electromagnetic modes of the cavities, i.e. $\omega_e=0$, and let us assume $N=2(2s+1)$ with $s$ an arbitrary integer number. Numerical computation of the eigenvalues $E$ of the linear system equation (\ref{matrice}) shows that for large $N$ the most localized mode, corresponding to the smallest value of the PR, is the one with energy $E=0$, and reads explicitly
\begin{equation}
a_n=0 \;, \;\; b_n=i^{n-1},  \;\;, \epsilon=- \frac{2 \rho}{\alpha}.
\end{equation}
 apart from a normalization factor. Note that this mode does not have any excitation in sublattice A, while the excitation is distributed in the emitter and uniformly in sublattice B. 
 The PR of this mode is given by
 \begin{equation}
 PR=\frac{\left[ \left(\frac{2 \rho}{\alpha} \right)^2 +N \right]^2 }{\left(\frac{2 \rho}{\alpha} \right)^4+N}.
 \end{equation}
Clearly, for large $N$ one has $PR \sim N$, i.e. this mode is not localized near the emitter and corresponds to a resonance state (the ratio between the amplitudes of excitation in the emitter and in the bath scales as $\sim \rho / \alpha$, diverging as $\alpha \rightarrow 0$, which is typical of a resonance state). This means that there is not any atom-field dark state in our system. 
\subsection*{Solution of equation (8)}
For $t<T^+$ we can neglect the sum over $\pm$ in equation  (8) of the main text, i.e.
\begin{eqnarray}
	\dot{\eps}(t)=-\tfrac{1}{2}\gamma_0\eps(t)-\sum_{n=1}^{\infty}\gamma_n^-
	\eps\!\left(t{-}nT^-\right)\Theta\!\left(t{-}nT^-\right),
\end{eqnarray}
with $\epsilon(0)=1$.
The above equation can be readily solved in Laplace domain. After introduction of the Laplace transform $\Tilde{\epsilon}(s)=\int_0^{\infty} dt \epsilon(t) \exp(-st)$, one obtains
\begin{eqnarray}
\Tilde{\eps}(s)=\left(s+\frac{\gamma_0}{2}-\frac{\gamma_0^-}{e^{s T^-}+1}
\right)^{-1},
\end{eqnarray}
where we assumed $N=2(2l+1)$ with $l$ integer. Expanding in power of $x=e^{sT^-}$ 
\small
\begin{eqnarray}
\Tilde{\eps}(s)=\frac{2}{2s{+}\gamma_0}+\sum_{n=1}^{\infty}4 \gamma_0^- (-1)^{n{+}1} e^{-n s T^-} \frac{(\gamma_0{-}2 \gamma_0^-{+}2 s)^{n{-}1}}{(\gamma_0{+}2 s)^{n{+}1}}.\nonumber
\end{eqnarray}
\normalsize
Returning into the time domain
\begin{eqnarray}
\eps(t)=e^{-\frac{\gamma_0t}{2}}{+}
\sum_{n=1}^\infty
(-1)^{n{+}1} \Theta(t{-}n T^-)\gamma_0^-(t{-}n T^-) {}_1F_{1}(n{+}1,2,-\frac{\gamma_0(t{-}n T^-)}{2}),
\end{eqnarray}
where ${}_1F_{1}$ is the confluent hypergeometric function.

	\begin{figure}[t!]
	\begin{center}
	     \includegraphics[width=0.9\linewidth]{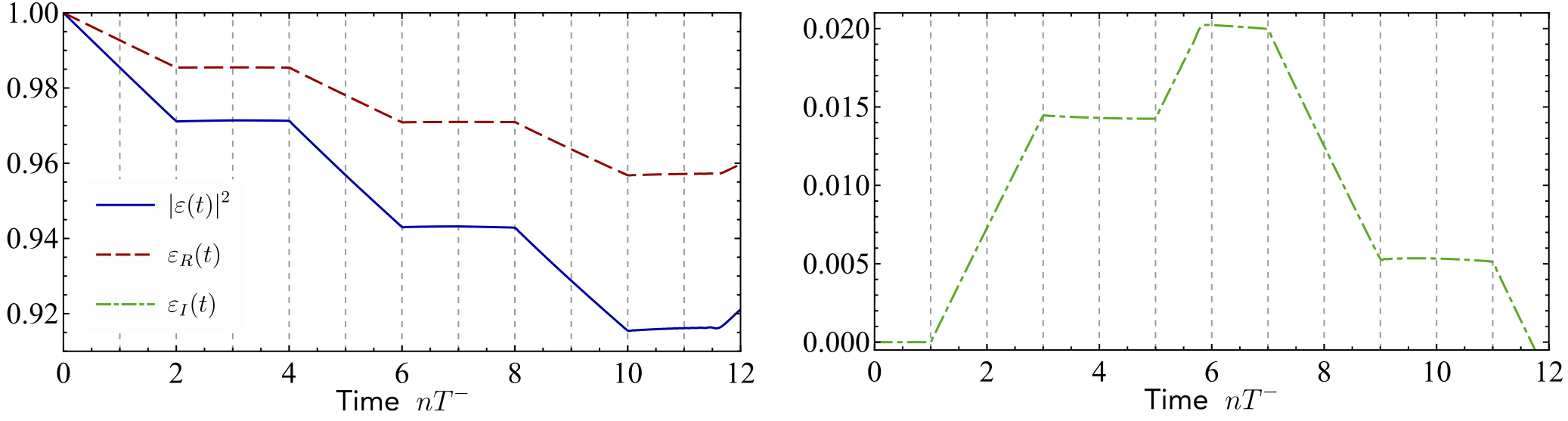}\\
	\end{center}
	\caption{Evolution of the squared excitation amplitude $|\eps(t)|^2$, solution of equation (6) in the main text and its real and imaginary parts $\eps_R(t)$, $\eps_I(t)$, for $N=501$, $J=\rho=1$ and $\alpha=0.01$. In this regime intermittent blockade of the decay is observed until the time $t \sim 12 T^-$.}
	\label{figSM1}
	\end{figure}

\subsection*{Decoherence dynamics for an odd number of sites in the ring} 

For the sake of completeness, let us discuss here the decay dynamics in a system with $N$ odd. As discussed in the main text, the dynamics of the coherences is accurately described by equation (8). In the case of an odd number of sites per sublattices,  $\gamma_n^\pm$ can take  imaginary values. As a consequence, and as shown in Fig. \ref{figSM1}, the intermittent blockade displays a doubled period which can be understood considering the approximate linearized solution for $t<T^+$. If we call $\eps_R(t)$ and $\eps_I(t)$ respectively the real and the imaginary part of $\eps(t)$, we have
\small
\begin{eqnarray}\eps_R(t){\sim}\begin{cases}
 1{-}\tfrac{1}{2}\gamma_0(t{-}2m T^-)\;&\text{for}\quad \;\;\;\;\;(4m)T^-{\leq}\;t\;{\leq} (4m{+}1)T^-\nonumber\\
  1{-}\tfrac{1}{2}\gamma_0(t{-}2m T^-)\; &\text{for}\quad(4m{+}1)T^-{\leq}\;t\;{\leq} (4m{+}2)T^-\nonumber\\
   1{-}\gamma_0(m+1) T^-\; &\text{for}\quad(4m{+}2)T^-{\leq}\;t\;{\leq} (4m{+}3)T^-\nonumber\\
    1{-}\gamma_0(m+1) T^-\; &\text{for}\quad(4m{+}3)T^-{\leq}\;t\;{\leq} (4m{+}4)T^-\nonumber\\
\end{cases}
\end{eqnarray}

\begin{eqnarray}\eps_I(t){\sim}\begin{cases}
 \pm \gamma_0 m T^-\quad &\text{for}\; \;\;\;\;\;(4m)T^-{\leq}\;t\;{\leq} (4m{+}1)T^-\nonumber\\
 \pm \tfrac{1}{2}\gamma_0(t{-}(2m{+}1)T^-)\; &\text{for}\quad(4m{+}1)T^-{\leq}\;t\;{\leq} (4m{+}2)T^-\nonumber\\
\pm \tfrac{1}{2}\gamma_0(t{-}(2m{+}1)T^-)\; &\text{for}\quad(4m{+}2)T^-{\leq}\;t\;{\leq} (4m{+}3)T^-\nonumber\\
   \pm \gamma_0 (m+1) T^-\; &\text{for}\quad(4m{+}3)T^-{\leq}\;t\;{\leq} (4m{+}4)T^-\nonumber\\
\end{cases}
\end{eqnarray}
\normalsize

\noindent for $m=0,1,2,3,\dots$, where for the imaginary part of the amplitude '$+$' corresponds to  $N=4s+1$  and  '$-$' to $N=2(2s+1)+1$, with $s$ integer. Thus, despite the short feedback period is still $T^-$, the dynamics displays four stages, in which the intermittency alternates between the real and imaginary parts of $\eps(t)$ in periods of $2T^-$. For the parameters considered in Fig. \ref{figSM1}, this linearized solution is accurate until $t\sim 6T^-$ for the imaginary part and $t\sim 12T^+$ for the real part, in which the first and second slow wavepackets interact with the two level system. Because of $\gamma^+_1$ is imaginary, the slow wavepacket needs two periods to affect the real part of the amplitude, and moreover, as in this early stage of the dynamics $|\eps_I(t)|\ll|\eps_R(t)|$ then $|\eps(t)|^2\approx \eps_R(t)^2$, and the intermittent blockade of the decay in $|\eps(t)|^2$ is significant until $t\sim 12T^-$ in contrast to the case studied in the main text in which it ceases at $t\sim 6 T^-$.





\section*{Acknowledgements}
We acknowledge the Spanish State Research Agency, through the Severo Ochoa and Mar\'ia de Maeztu Program for Centers and Units of Excellence in R\&D (MDM-2017-0711) and through the  QUARESC project (PID2019-109094GB-C21 and -C22/ AEI / 10.13039/501100011033); we also acknowledge CSIC Research Platform PTI-001, the PIE project 202050E098,  the QUAREC project funded by CAIB, and  funding from CAIB PhD program.
 GLG is funded by the Spanish  Ministerio de Educaci\'on y Formaci\'on Profesional / Ministerio de Universidades   and  co-funded by the University of the Balearic Islands through the Beatriz Galindo program  (BG20/00085).  Sa. L. acknowledges  support from MIUR through project PRIN Project 2017SRN-BRK QUSHIP and hospitality from IFISC under the ``professors convidats” UIB program. Sa.L. thanks F. Ciccarello for useful discussions.


\section*{Author contributions statement}
Sa. L., St. L. G. L. G. and R. Z. conceived the setup, Sa. L. and A. C. developed the analytical calculations, Sa.L. and G. L. G. performed the numerical simulations. All authors analysed the results and contributed to the writing of the manuscript.

\section*{Additional information}

\textbf{Competing interests} The authors declare no competing interests.



\end{document}